\documentclass{PoS}

\usepackage{amsmath,amssymb,amsbsy}
\usepackage{mathrsfs,stmaryrd,eucal}

\newcommand{\be}{\begin{equation}}
\newcommand{\ee}{\end{equation}}
\newcommand{\ba}{\begin{eqnarray}}
\newcommand{\ea}{\end{eqnarray}}

\newcommand{\f}{\frac}

\newcommand{\rd}{{\rm d}}
\newcommand{\vp}{\varphi}
\newcommand{\pa}{\partial}

\title{General Relativistic Rotation Curves in a Post-Newtonian Light}

\ShortTitle{General Relativistic Rotation Curves in a Post-Newtonian Light}

\author{\speaker{Aleksandar Raki\'c}\\
        Institut f\"ur Theoretische Physik und Astrophysik, Universit\"at W\"urzburg, D-97074 W\"urzburg, Germany\\
        E-mail: \email{rakic\_at\_astro.uni-wuerzburg.de}}

\author{Dominik J. Schwarz\\
        Fakult\"at f\"ur Physik, Universit\"at Bielefeld, D-33501 Bielefeld, Germany\\
        E-mail: \email{dschwarz\_at\_physik.uni-bielefeld.de}}

\abstract{
The missing of a Keplerian fall-off in the observed galaxy rotation curves represents classical evidence for the existence of dark matter on galactic scales. There has been some recent activity concerning the potential of modelling galactic systems with the help of general relativity. This was motivated by claims that by the use of full general relativity dark matter could be made superfluous. Here we focus on possible axisymmetric and stationary solutions of Einstein's equations with rotating dust. After a short review of the current debate we pursue the idea of approaching such relativistic models in a Newtonian language. We analyse rigidly as well as differentially rotating Newtonian and Post-Newtonian spacetimes and find that it is necessary to incorporate a Post-Newtonian term in order to make physical sense.
}

\FullConference{Identification of dark matter 2008\\
          August 18-22, 2008\\
          Stockholm, Sweden}

\begin{document}

\section{Motivation}
%A huge amount of current observations in cosmology ask for a parameterisation in terms of dark matter in order to make sense within the standard model. The concordance model that enfolds observations of the CMB as well as of the large scale structure currently charges a value of $\Omega_{\rm dm}\simeq0.23$ \cite{lambda}. New ans\"atze ranging from modifications of the standard theory of particles to alternative gravity theories are constantly being thrown at the problem. Nonetheless the identification of dark matter is still at best incomplete.

The evidence for dark matter stems from various observations at various scales but it is not necessary that the effects have all the same origin. This paper is motivated by the galactic dark matter problem: in contrast\footnote{In Newtonian gravity a flat rotation curve is possible, see Mestel's disk \cite{mestel}. But the price is an infinite total mass.} to the expectation from Newtonian dynamics the tangential velocities of spirals are found to be approximately flat and not decreasing up to large radii.

The standard resolution lies in postulating a large halo of dark matter around galaxies. In contrast to this Cooperstock and Tieu (CT) recently suggested that the proper modelling of a galaxy through full general relativity (GR) explains the findings without dark matter \cite{cooperst}. Their galaxy model is a stationary axisymmetric (comoving dust) solution based upon the following metric
\be
 \rd s^2 = e^{\nu - w} \left( \rd r^2 + \rd z^2 \right) + r^2 e^{-w} \rd \vp^2 - e^w \left( \rd t + N \rd \vp \right)^2 \,.
\label{eq:ctmetric2}
\ee
Even though fields and velocities in the galaxy are small, according to CT, the Newtonian approximation would not be valid for a galactic system that is under the influence of gravity alone because of intrinsic non-linear contributions of non-negligible size from GR. The GR model would then be able to reproduce realistic density profiles after fitting flat rotation curves. The according total galactic mass comes out of modest value, c.f.~\cite{cooperst}, that is smaller than the newtonianly required value and larger than values from Modified Newtonian Dynamics (MOND).

By now various authors have pointed to problems with the consistency of the CT model. In \cite{korzynsk} the Komar mass within the CT model is analysed and it is found that the energy momentum tensor becomes ill-behaved at $z=0$, revealing the existence of an additional exotic matter source. Thus, in the CT model, the location of exotic matter would only be transfered to the galactic plane, a finding also supported in \cite{letelier}. In \cite{garfinkle} it is argued that Post-Newtonian corrections should already enfold non-linear effects if they are present and that corrections should intrinsically be small in the weak field limit. In \cite{cross} an inconsistency between the CT comoving frame and the condition of differential rotation is revealed; if the inconsistency is removed the rotation curve became ordinarily Keplerian. It is argued in \cite{fuchs} that the CT model gives wrong predictions for the lateral density profile of the Milky Way and for its local density. On the other hand, in \cite{menzies} it is found that the CT model seems to hide infinite mass at large distances. Generally it is shown in \cite{zingg} that interior stationary and axisymmetric solutions - like CT - cannot be continued into according exterior solutions in order to become global solutions without evoking singularities.

It is interesting to note that viable disk models for galactic systems with differential rotation do not exist in GR by now. For the case of rigid rotation \cite{neuge} an exact solution is known which has been obtained via the inverse scattering method; already its formulation is technically rather elaborate. Though problematic, the CT hypothesis has initiated further alternative ideas on galactic rotation curves. Using a similar GR model in \cite{balasin} it is found that the amount of dark matter needed is reduced by 30\%. In \cite{coimbra} an axisymmetric disk solution of the Einstein equations in six dimensions is constructed. The modification of gravity via extra dimensions then leads to flat rotation curves if orbits within the disk are stable.

Here we want to compare the CT model with a rotating (Post-) Newtonian model. For this we first have to ask some general questions on axially symmetric and stationary solutions in GR.

\section{Rotating (Post-) Newtonian Spacetimes}%Axisymmetric and Stationary Spacetimes}
In notation $(x^0,x^1,x^2,x^3)=(t,r,\vp,z)$ and signature $(-,+,+,+)$ the most general axisymmetric $(\partial_\vp g_{\mu \nu}=0)$ and stationary $(\partial_t g_{\mu \nu}=0)$ spacetime in GR with the four free metric functions $U(r,z),k(r,z),W(r,z),A(r,z)$ is named after Lewis and Papapetrou (LP) and reads \cite{stephani}
\be
 \rd s^2 = e^{-2U} \left[ e^{2k}(\rd r^2 + \rd z^2) + W^2 \rd\varphi^2 \right] - e^{2U} ( \rd t + A \rd\varphi)^2 \,.
\label{eq:lpmatricfin}
\ee
%with the free metric functions $U,k,W,A$ being functions of only $r$ and $z\,$.
We can simplify (\ref{eq:lpmatricfin}) a bit more, but only under crucial assumptions. If and only if the metric function $W$ is harmonic it can be transformed to $W=r\,$. Consider a complex coordinate transformation $f(r+iz) = W + iV$ introducing an additional potential $V\,$. Then we have from $\rho \equiv W(r,z)$ and $h \equiv V(r,z)$ the differentials $\,\rd \rho = \f{\pa W}{\pa r} \rd r + \f{\pa W}{\pa z} \rd z\,$ and $\,\rd h = \f{\pa V}{\pa r} \rd r + \f{\pa V}{\pa z} \rd z\,$. After inserting into (\ref{eq:lpmatricfin}), written in terms of $\rho,h\,$, and requiring formal invariance as compared to the original metric we see that the mixing terms have to vanish. That is exactly provided by the Cauchy-Riemann equations for $W$ and $V$: $\,\frac{\partial W}{\partial r} = \frac{\partial V}{\partial z}\,$ and $\,\frac{\partial W}{\partial z} = - \frac{\partial V}{\partial r}\,$. Moreover, with the help of these equations, we see that the coefficients of $\rd r^2$ and $\rd z^2$ can be combined and so we reobtain (\ref{eq:lpmatricfin}). 

Thus we have shown that it is possible to simplify the general LP form (\ref{eq:lpmatricfin}) by allowing $W = r\,$, which is only possible if $W$ (and $V$) be harmonic\footnote{This condition for $W$ holds for exterior (vacuum) solutions that are stationary and axisymmetric, c.f.~\cite{islam,wald}.}: $\Delta^{(2)} W = 0\,$. Upon this constraint we can write down the LP metric in isotropic coordinates (or Weyl gauge)
\be
 \rd s^2 = e^{-2U} \left[ e^{2k}(\rd r^2 + \rd z^2) + r^2 \rd\varphi^2 \right] - e^{2U} ( \rd t + A \rd\varphi)^2 \,.
\label{eq:lpweylg}
\ee
The spacetime applied in the CT model was (\ref{eq:ctmetric2}). Obviously, the CT metric does not belong to the class of the most general stationary and axisymmetric spacetimes; it only belongs to the subclass of LP solutions in the Weyl gauge, and is therefore less general.

Now we ask what solution could potentially be a Newtonian counterpart to the CT model (\ref{eq:ctmetric2}). The \lq Newtonian approximation\rq \cite{mtw} that is the metric that reproduces Newtonian physics is
\be
  \rd s^2 = -(1 + 2\phi)\rd t^2 + \rd r^2 + r^2 \rd \vp^2 + \rd z^2 \,,
\label{eq:naivenewton}
\ee
where $\phi=\phi(r,z)$ is the Newtonian gravitational potential. For simplicity, we start with rigid rotation in (\ref{eq:naivenewton}) via $\varphi = \varphi^\prime - \omega \, t$ and $\omega = $const. The exact result can be brought to the form
\be
 \rd s^2 = (\rd r^2 + \rd z^2) + \frac{1 + 2\phi}{(1 + 2\phi - \omega^2 r^2)} r^2 \rd\varphi^2 - (1 + 2\phi - \omega^2 r^2) \left[\rd t + \frac{r^2 \omega}{(1 + 2\phi - \omega^2 r^2)} \, \rd\varphi \right]^2 \,.
\label{eq:newstifexa}
\ee
In this form we can directly compare it with the LP metric in Weyl gauge (\ref{eq:lpweylg}), and we notice a discrepancy at linear order in $\phi\,$, looking at the $\rd \vp^2$ term. Interestingly, the rigidly rotated Newton metric (\ref{eq:newstifexa}) is not consistent with the Weyl subclass of the LP solution (\ref{eq:lpweylg}).

One could now ask whether the situation might be easily cured with the help of a coordinate transformation. We show that this is not possible. Above we have derived the exact conditions under which the general and the isotropic LP metric can be transformed into each other, that is the function $W$ must fulfil the two-dimensional Laplace equation $\Delta^{(2)} W = 0\,$. In the case of (\ref{eq:newstifexa}) we identify $W = r \sqrt{1 + 2 \phi}\,$. Expanding to linear order and applying the Laplacian yields
\be
 \Delta^{(2)} W = r \Delta^{(3)}\phi + \phi_{,r} = 4\pi G\rho r + \phi_{,r}
\label{eq:lapl2W}
\ee
after using the Poisson equation. We will now show that $\Delta^{(2)} W$ does not vanish in general.

Given the problem of solving the Laplace equation with boundary conditions for a disk-like distribution of matter, the solution for the potential can be obtained \cite{binney} via separation of variables: $\phi(r,z) = \int_0^\infty S(k) J_0(kr) e^{-k |z|} \rd k\,$. A surface mass density $\Sigma(r)$ is then characterised by its Hankel transform $S(k) = - 2 \pi G \int_0^\infty J_0(kr) \Sigma(r) r \rd r\,$. We use these expressions for the evaluation of (\ref{eq:lapl2W}).

%\footnote{The flat rotation curve in the Mestel model is obtained from the Hankel transform of (\ref{eq:mestelsurfden}), inserted into the formula for the rotation curve: $v^2(r)_{\rm Mes} = r (\pa \phi/\pa r)_{z=0} = 2\pi G\Sigma_0 r_0\,$.}

Case (I) $z\neq 0\,$. Outside the disk the Newtonian potential fulfils the Laplace equation, such that (\ref{eq:lapl2W}) becomes $\Delta^{(2)} W = - \int_0^\infty S(k) J_1(kr) k e^{-k |z|} \rd k$ which will not vanish in general. We demonstrate this e.g.~with the Mestel model \cite{mestel} which is characterised by a surface mass density that falls off inversely with the distance $\Sigma(r) = \Sigma_0 r_0/r\,$ and Hankel-transforms as $ S(k) = - 2 \pi G \Sigma_0 r_0/k $. Using this we can integrate and obtain $\Delta^{(2)} W = 2 \pi G \Sigma_0 r_0 \left( \f{1}{r} - \f{|z|}{r\sqrt{r^2+z^2}} \right)$ at $z \neq 0\,$.

Case (II) $z = 0\,$. We want to show that (\ref{eq:lapl2W}) is non-zero also here. Let us assume the contrary and see what happens. If we assume that $\Delta^{(2)} W = 0$ was true then equation (\ref{eq:lapl2W}) gives an identity. This we integrate over $z$ for some $\varepsilon > 0$ and then revoke the operation with the appropriate limit, $\, -4 \pi G r \lim_{\varepsilon\to 0} \int_{- \varepsilon}^\varepsilon \delta(z) \Sigma(r) \rd z = \lim_{\varepsilon\to 0} \int_{- \varepsilon}^\varepsilon \int_0^\infty S(k) J_1(kr) k e^{-k|z|} \rd k \rd z\,$. Since the exponential term on the right hand side serves as a damping factor, the modulus of the integrand will reach its maximum at $z=0$. Thus, as an upper estimate, we can set the integrand of the right hand side to be constant in $z$ and therefore the integration and limit procedure give zero. Then, for all other $z$ the expression will be zero more than ever and we have $4 \pi G \Sigma(r) r = 0\,$: this will not hold generally for any realistic model, and thus $\Delta^{(2)} W(r,z) = 0$ is not true at the surface $z=0$ either.

\emph{Notably, the classical Newton metric (\ref{eq:naivenewton}) cannot be made compatible with the LP metric in Weyl form and is thus also not compatible with the CT model. Additionally, as all stationary and axisymmetric vacuum solutions belong to the Weyl class} \cite{wald}, \emph{it follows that such a galaxy does not have a rotating classically Newtonian exterior solution.} We go one step further and consider the Post-Newtonian\footnote{Sometimes, this metric with $\psi=\phi$ is referred to as the \lq Newton metric\rq\ in the literature. The reason for this nomenclature might be that the order of magnitude of the coefficient of the spatial part and the order of the Newtonian correction are the same. However, conceptually this makes a considerable difference. In classical Newton Gravity there exists no curvature of space, the three-space is always euclidian. This is exactly reflected in the Newton metric (\ref{eq:naivenewton}) and therefore we refer to (\ref{eq:ppnmetric}) as the PN approximation; for an extensive discussion see e.g.~chp.~39 of \cite{mtw}.} (PN) metric with an additional PN potential $\psi=\psi(r,z)\,$,
\be
 \rd s^2 = -(1 + 2\phi)\rd t^2 + (1 - 2\psi) (\rd r^2 + r^2 \rd \vp^2 + \rd z^2) \,.
\label{eq:ppnmetric}
\ee
To include one more complication we consider the \emph{differential rotation} of this metric via $\omega = \omega(r,z)\,$ and arrive at a lengthy expression. Unfortunately, the resulting metric exhibits direct time dependence in some coefficients and is therefore only reasonable in a strictly local sense. To preserve stationarity we approximate it by allowing for only small time intervals or equivalently for small angles of rotation. We find that for the simplest PN case, i.e.~$\phi=\psi\,$, the differentially (and the rigidly) rotated metric is consistent - neglecting terms $\mathcal{O}(\phi^2)$ - with the Weyl form (\ref{eq:lpweylg}).

\section{Summary and Conclusion}
Up to day, there does not exist an intuitive and applicable GR solution which could help to realistically model a galactic system - except for the complicated solution \cite{neuge} with stiff rotation. CT have proposed a solution but as we have seen arguments from various directions have revealed unphysical features in the model. We have shown that the CT solution does not belong to the class of the most general axially symmetric and stationary solutions, the LP class. Therefore the CT solution is less general and this restriction might be connected with the problems of the model.

A more intuitive understanding of the CT model and its breakdown would be instructive. For this we approached the CT model by Newtonian means. We found the interesting result that the classical Newton metric after rotation does not lie in the same class as the CT metric but is more general and cannot be simplified. Only the incorporation of a PN potential gives rise - after differential or rigid rotation - to equivalent metrics. Thus the proper weak field limit of such a galaxy must be Post-Newtonian. A complete analysis of the resulting dynamics via the Arnowitt-Deser-Misner formalism is the concluding step and is going to be published elsewhere.

%This work was supported by the DFG under grants GRK 881 and GRK 1147.

%Thus the PN metric after rigid rotation belongs to the class of isotropic (Weyl) axisymmetric and stationary solutions, if $\phi=\psi$, whereas the rotated Newton metric (\ref{eq:newstifexa}) does not allow for that simplification.

%is thus also perfectly compatible with the Weyl gauge of the LP metric under the condition $\phi=\psi\,$.

\end{document}